\documentstyle[12pt]{article}
\def\etal{et al.}
\def\plb#1{Phys. Lett. B {#1}}
\def\npb#1{Nucl. Phys. B { #1}}
\def\npbps#1{Nucl. Phys. B (Proc. Suppl.) { #1}}

\newcommand{\beq}{\begin{equation}}
\newcommand{\eeq}{\end{equation}}
\newcommand{\beqn}{\begin{eqnarray}}
\newcommand{\eeqn}{\end{eqnarray}}
\newcommand{\bea}{\begin{eqnarray}}
\newcommand{\eea}{\end{eqnarray}}

\begin{document}
\pagestyle{empty} 
\vspace{-0.6in}
\begin{flushright}
ROME1-1206/98 \\
TUM-HEP-312/98 \\
\end{flushright}
\vskip 1cm
% Definition of title page:
\centerline{\large{\bf{MODEL INDEPENDENT DETERMINATION OF}}}
\centerline{\large{\bf{THE LIGHT-CONE WAVE FUNCTIONS}}}
\centerline{\large{\bf{FOR EXCLUSIVE PROCESSES}}}
\vskip 1.4cm
\centerline{U.~Aglietti$^1$, M.~Ciuchini$^{2}$, G.~Corb\`o$^1$,}
\centerline{E.~Franco$^1$, G. Martinelli$^{1}$, L.~Silvestrini$^3$}
\vskip  0.4cm
\centerline{\small $^1$ Dipartimento di Fisica, 
Universit\`a  di Roma ``La Sapienza" 
and INFN,}
\centerline{\small Sezione di Roma, P.le A. Moro 2, I-00185 Rome, Italy.}
\centerline{$^2$ Dipartimento di Fisica, Universit\`a  di Roma Tre and INFN,}
\centerline{Sezione di Roma Tre, Via della Vasca Navale 84, I-00146
 Rome, Italy.}
\centerline{\small $^3$ Physik Department, Technische Universit\"at M\"unchen,}
\centerline{\small D-85748 Garching, Germany.}
\vskip 1.5cm
\abstract{
We present  a   method to compute, by numerical  simulations of 
lattice QCD, the light-cone wave functions which enter  exclusive
processes at large momentum transfer, such as  electromagnetic elastic
scatterings,  or    exclusive semi-leptonic decays as
$B \to \pi$ ($B \to \rho$)  and   radiative decays as $B \to K^* \gamma$.  
The method is based  on first principles and does not require 
any model assumption.}
\eject
\pagestyle{empty}\clearpage
\setcounter{page}{1}
\pagestyle{plain}
\newpage 
\pagestyle{plain} \setcounter{page}{1}
%\section{Introduction}
%\label{sec:intro}
In this paper  we propose a method  to compute, by numerical simulations
on the lattice, the light-cone
wave functions which allow to predict the form factors 
relevant in many exclusive processes, such as 
  electromagnetic elastic scattering  at large momentum
transfer,  or    exclusive semi-leptonic decays  as $B \to \pi$ ($B \to \rho$) 
and   $B \to K^* \gamma$ decays~\cite{general}--\cite{ball}.
Our approach allows to compute the form factors in
exclusive semi-leptonic $B$ decays at small values of the invariant mass of the
lepton pair, $q^2$, a region which is not accessible by standard lattice
techniques, which are confined in the region 
$q^2 \sim q^2_{max}$~\cite{lanl:semilept-lat94}--\cite{ukqcd:hlfits}.
We also show that, from suitable combinations
of lattice correlation functions, it is possible to determine
the full light-cone wave function, denoted generically  as $\Phi$,
 and not only  its  moments,
 as done in the past~\cite{sac1}. The possibility of computing
directly  $\Phi$ is a significative advantage:  
higher  moments of the light-cone wave functions, being related  
to higher dimension operators, are  in general
afflicted by power divergencies when a hard cutoff~\footnote{  In the
lattice regularization  the hard
cutoff  is given by the inverse lattice spacing $1/a$.} is used (or 
renormalon ambiguities in dimensional regularizations). This makes 
very  problematic the  definition of the renormalized operators, i.e. those
which have  finite matrix 
elements when the cutoff is removed~\cite{smrenormalon}. With our method, instead, 
no renormalization is needed as  we get directly $\Phi$ from
the appropriate correlation functions. 
The technique described below  strictly follows a similar proposal recently made 
to determine the shape function for  inclusive heavy-hadron decays~\cite{ua}.
\vskip 0.7 cm
\par We now explain how the method works.
The light-cone wave functions are universal quantities, i.e. quantities independent
from the process at hand (electromagnetic elastic scattering 
at large  momentum transfer, exclusive $B$ decays, etc.).
Thus, in order  to illustrate our proposal,
we  start by considering a very simple prototype which enters, for example,
the elastic $\gamma^*(q) + \pi \to \pi$ scattering~\cite{general} and 
$B \to \pi$ semileptonic decays~\cite{ruckl} 
\beq  F^\mu( \vec p_\pi,q) \equiv i \int d^4x \ e^{i q \cdot x} 
\langle \pi^-(\vec p_\pi) \vert \bar d(x) \gamma^\mu \gamma_5 S(x;0) u(0) 
\vert 0 \rangle \, \label{eq:proto} \eeq
where, for simplicity, and without loss of generality, we have taken $S(x;0)$
to be the scalar Feynman propagator, satisfying the differential equation
\beq  - D^2 S(x;0) =\delta^4(x-0) \, . \label{eq:full} \eeq 
With this choice the multilocal operator appearing in eq.~(\ref{eq:proto})
is gauge-invariant.
\par $F^\mu(\vec p_\pi,q)$ can be written as 
\beq \label{eq:fmue} F^\mu(\vec p_\pi,q) = 
i \int d^4x \ 
\langle \pi^-(\vec p_\pi) \vert \bar d(x) \gamma^\mu \gamma_5 S_Q(x;0) u(0) 
\vert 0 \rangle  \ , \eeq 
where  $S_Q(x;0)\equiv e^{i q \cdot x} S(x;0)$. For simplicity we 
consider first the case of a pion with  a small momentum (at rest), i.e. 
$\vert \vec p_\pi \vert \ll \vert \vec q \vert$ ($\vec p_\pi=0$).
For $-q^2 = Q^2 \gg p_\pi^2 = M_\pi^2$,  we can separate  the large frequency 
modes ($\sim Q=\sqrt{Q^2} $) from the low energy modes
($\sim \Lambda_{QCD}$) and expand   $S_Q(x;0)$ in powers of 
$\Lambda_{QCD}/Q$   
\beq \label{eq:expaSQ} S_Q(x;0) \simeq
\left( \frac{1}{ - Q^2 + 2 i q \cdot
D + i \epsilon }\right)_{(x;0)} \ . \eeq
In   eq.~(\ref{eq:expaSQ}) we  kept only the leading terms
of the expansion in powers of $1/Q$, and those which become leading 
when $u=Q^2/(2 p_\pi \cdot q )\sim 1$.
Using  eq.~(\ref{eq:expaSQ}) one finds
\beq \label{eq:deff+} F^\mu(\vec p_\pi,q) = 
f_\pi p_\pi^\mu \int_0^1 du \ \frac{\Phi_\pi(u)}{-Q^2 + 2 u q \cdot p_\pi + i \epsilon}
 \eeq 
where the light-cone wave  function $\Phi_\pi(u)$ is defined through the relations
\beqn 
\langle  \pi^-(\vec p_\pi) \vert \bar d(0)\gamma^\mu \gamma_5
(i D^{\mu_1}) \dots
(i D^{\mu_n}) u(0)  \vert 0 \rangle &=& -i f_\pi {\cal M}_{n} p_\pi^\mu p_\pi^{\mu_1}
\dots p_\pi^{\mu_n} \nonumber \\  &+& i {\cal B}_n \delta^{\mu \mu_1}  
p_\pi^{\mu_2} \dots p_\pi^{\mu_n} +\dots \, , \label{eq: momentD} \eeqn
with 
\beq \label{eq:phidef} {\cal M}_n = \int_0^1 du \ u^n \Phi_\pi(u) \, . \eeq
In eq.~(\ref{eq: momentD}), the contributions proportional to ${\cal B}_n$
are suppressed by higher powers of $1/Q$ in all the relevant 
kinematical region.
\par We now consider the  Fourier transform of $F^\mu(\vec p_\pi,q)$
defined as
\beq \label{eq:tmnt} F^\mu(t, \vec p_\pi, \vec q)= \int \frac{d q_0 }{2 \pi}
e^{- i q_0 t} F^\mu(\vec p_\pi,q) \ .\eeq
For $t \ge 0$, by closing the contour of  integration over $q_{0}$
below the real axis,  we find
\bea \label{eq:tmntn} F^\mu(t, \vec p_\pi, \vec q) = 
-i f_\pi p_\pi^\mu \int_0^1 du \ 
\Phi_\pi(u)  \frac{e^{- i (- u E_\pi  + 
\sqrt{\vec q^2 + u^2 E_\pi^2+ 2 u \vec q \cdot \vec p_\pi } ) t } }
{2 \sqrt{\vec q^2 + u^2 E_\pi^2+ 2 u \vec q \cdot \vec p_\pi } } \ .
\eea 
If,  for consistency with the order at which we are working, we neglect
the terms of ${\cal O}(M_\pi^2/Q^2)$ and $ {\cal O}(\vec p^2_\pi  / \vec
q^2 )$, we get
\bea  \label{eq:tmntne} F^\mu(t, \vec p_\pi, \vec q) 
\simeq  - i f_\pi p_\pi^\mu \int_0^1 du \ 
\Phi_\pi(u)  \frac{e^{- i  (- u E_\pi  + 
\sqrt{\vec q^2+ 2 u \vec q \cdot \vec p_\pi} ) t } }
{2 \sqrt{\vec q^2+ 2 u \vec q \cdot\vec p_\pi}}\ .
\eea
It is convenient to write  eq.~(\ref{eq:tmntn}) as follows
\bea \label{eq:tmntg} F^\mu(t, \vec p_\pi, \vec q) = 
-i f_\pi p_\pi^\mu \int_0^1 du \ 
\Phi_\pi(u)  \frac{e^{- i u (- E_\pi(\vec p_\pi)  +  E_\pi(\vec p_\pi+ \vec q
/ u ) )
 t } }{ 2 u E_\pi(\vec p_\pi+ \vec q / u ) } \ , \eea 
where $E_\pi(\vec p) =\sqrt{\vec p^2 + M_\pi^2}$.  In the latter form,  one
recognizes  that  eq.~(\ref{eq:tmntn})
is valid for arbitrary pion momenta. 
In the general case, eqs.~(\ref{eq:tmnt})--(\ref{eq:tmntg}) can be derived from
the following expression
\beq \label{eq:deff+g} F^\mu(\vec p_\pi,q) = 
f_\pi p_\pi^\mu \int_0^1 du \ \frac{\Phi_\pi(u)}{W^2 + 2  (q +p_\pi) 
\cdot k + i \epsilon} \ ,
 \eeq 
where $W^2=(q+p_\pi)^2=Q^2(1-x)/x$ is the squared invariant  mass of the
recoiling hadron system and  $k= u p_\pi - p_\pi$ is the residual momentum 
 of the struck  parton 
($\langle D^{\mu_1} \dots D^{\mu_n} \rangle \sim k^{\mu_1} \dots k^{\mu_n}
= {\cal O}(\Lambda_{QCD}^n)$).   The expression above 
shows that different dynamics  which  occur depending on the value of $x$.
At small $x$, we can  retain only the $W^2$ term in the denominator and
the amplitude becomes proportional to the lowest moment of $\Phi_\pi$ which
is unity;
when $W^2 \sim  Q \Lambda_{QCD}$, all the moments become comparable and 
we need  the full wave function $\Phi_\pi$; finally in 
the region where $W^2 \sim
\Lambda^2_{QCD}$,  the  light-cone approach fails. This is the analog
of what happens in inclusive decays at the end point of the energy 
spectrum~\cite{ua}.
\par 
It is straightforward to make the analytic continuation of the above expression to the
Euclidean space-time which is used in numerical simulations 
\bea \label{eq:tmnte} F^\mu(t, \vec p_\pi, \vec q) &\equiv&
-i f_\pi p_\pi^\mu   F(t, \vec p_\pi, \vec q) \nonumber \\ &=&
-i f_\pi p_\pi^\mu \int_0^1 du \ 
\Phi_\pi(u)  \frac{e^{-  u (- E_\pi(\vec p_\pi)  +  E_\pi(\vec p_\pi+ \vec q
/ u ) )
 t } }{ 2 u E_\pi(\vec p_\pi+ \vec q / u ) } \ , \eea 
By studying the time-dependence of $F^\mu(t, \vec p_\pi, \vec q)$ at several values 
of $\vec q$ and $\vec p_\pi$, we can unfold the integral above and
and extract the light-cone  wave function $\Phi_\pi$.
\vskip 0.7 cm
We give  below the practical recipe to implement the calculation of 
$F^\mu(t, \vec p_\pi, \vec q)$ in lattice simulations.
From the definition of $F^\mu(\vec p_\pi,q)$ in eq.~(\ref{eq:proto}),
one has 
\beq  F^\mu(t, \vec p_\pi, \vec q) = i  \int d^3 x \ 
e^{-i \vec q \cdot \vec x} \ \langle \pi(\vec p_\pi) \vert \bar d(\vec x ,t) \gamma^\mu \gamma_5
S(\vec x,t ; \vec 0, 0 ) u(0) \vert 0 \rangle \label{eq:promo} \ .\eeq
The above amplitude can be extracted directly from a suitable ratio
of lattice three- and two-point correlation functions 
\beq R(t, \vec p_\pi, \vec q )= \lim_{t_f \to \infty}
e^{- E_\pi t}
\frac{G^\mu_3( t_f, t , \vec p_\pi, \vec q )}{G^\mu_2(t_f, \vec p_\pi) } 
\, , \label{eq:rmu}  \eeq
where $E_\pi =\sqrt{M_\pi^2 +\vec p_\pi^2}$;
\beq G^\mu_3( t_f, t , \vec p_\pi, \vec q ) = 
\int d^3 x \  e^{i \vec q \cdot \vec x } \ 
\langle 0 \vert \Pi_{\vec p_\pi}(t_f) \bar d(\vec x ,t) \gamma^\mu \gamma_5
S(\vec x,t ;\vec 0, 0 ) u(0) \vert 0 \rangle \label{eq:g3} \ \eeq
and 
\beq   G^\mu_2(t_f, \vec p_\pi) =
\langle 0 \vert  \Pi_{\vec p_\pi}(t_f)  A^{\mu \dagger}_{\vec p_\pi}(0)
\vert 0 \rangle \  .
\label{eq:g2} \eeq
$\Pi_{\vec p_\pi}(t)$ and $A^{\mu \dagger}_{\vec p_\pi}(t)$
are  the pion interpolating field and the axial  current
($A^\mu(x) = \bar u(x) \gamma^\mu \gamma_5 d(x)$)
with definite spatial momentum $\vec p_\pi$ 
\beq
\Pi_{\vec p_\pi}(t)=\int d^3x e^{ i\vec p_\pi\cdot\vec x}
\partial_\mu A^\mu(\vec x,t)\ , \quad 
A^\mu_{\vec p_\pi}(t)=\int d^3x e^{ i\vec p_\pi\cdot\vec x}
 A^\mu(\vec x,t) \  . \eeq
\par  In the Euclidean,
using the transfer matrix formalism, we have
\beq 
\Pi_{\vec p_\pi}(\vec x , t)=e^{\hat H t} \Pi_{\vec p_\pi}(\vec x) e^{-\hat H t} \ ,
\eeq
so that the correlation functions have an exponential dependence on the energy
of the external states. 
This implies that, in the limit $t_f\to\infty$, the
lightest  state, corresponding to a pion, dominates the
correlation functions (\ref{eq:g3}) and (\ref{eq:g2}), since all 
higher-energy   states are exponentially suppressed.
In this limit, with $t > 0 $, we then obtain
\beq \label{eq:rwn} R(t, \vec p_\pi, \vec q ) \to 
F(t, \vec p_\pi, \vec q) \, , \eeq
which is the desired quantity.
\par It is obvious that the same technique can be used to compute all the
light-cone functions which have been introduced in the literature.
In the case of the pion, for example, those  denoted as
$\Phi_p$ and $\Phi_\sigma$ in ref.~\cite{ruckl}  correspond to the non-local
amplitude of eq.~(\ref{eq:proto}), where $\gamma^\mu \gamma_5$ is replaced
by $i \gamma_5$ and $\sigma_{\mu\nu}$ respectively. 
The same holds true, taking into account  all the complications entailed
by the polarization of the physical states, in the case of  the $\Phi$s
which are relevant for  the $\rho$~\cite{ball}, the $K^*$ meson~\cite{ali} 
and the  nucleon~\cite{cernyak} cases.
\par In practical calculations  one may use either 
the scalar propagator defined in eq.~(\ref{eq:full}) 
or  the approximate  propagator introduced in  eq.~(\ref{eq:expaSQ}). In the general
case, the latter can be written as
\beq \label{eq:expaSQp} S_Q(x;0) =
\left( \frac{1}{ (q+p_\pi)^2 + 2 i (q +p_\pi) \cdot
D + i \epsilon }\right)_{(x;0)} \ . \eeq
$S_Q(x)$ can be written as 
\beq   S_Q(x) =\frac{e^{+ i Q   x_+ /2}}{Q}  S_{\scriptscriptstyle
LEET}(x) \, , \eeq
where $x_+ = n\cdot x $, with $n^\mu= (q^\mu + p^\mu_\pi)/Q$,
and $S_{\scriptscriptstyle LEET}(x)$ is
the light-cone   propagator of the Large
Energy Effective Theory (LEET)~\cite{grin}, which satisfies the equation
\beq 2 i D_+ S_{\scriptscriptstyle LEET}(x) =\delta^4(x) \, .\eeq 
$n^\mu$ is the appropriate vector which becomes light-like in the elastic region,
$n^2 =  W^2/Q^2 \sim  \Lambda_{QCD}/Q$ when $W \sim  Q \Lambda_{QCD}$.
Thus, the extraction of  the  $\Phi$s from $S_Q(x)$ is 
equivalent to the use of the LEET.  
Note that the calculation of the   physical light-cone function
$\Phi_\pi$ using the full propagator of  eq.~(\ref{eq:full})
does not need  any renormalization (for $\Phi_p$ and $\Phi_\sigma$,
the same renormalization constants of the operators $\bar \psi \gamma_5
\psi$ and $\bar \psi \sigma^{\mu\nu} \psi$ must be applied). 
The same calculation 
using the $S_{\scriptscriptstyle LEET}$   requires, instead
an overall  (further) logarithmic renormalization of the 
amplitude~\cite{koster}, which can be computed in lattice perturbation theory.
The ultraviolet divergences of the LEET correspond perturbatively
to  infrared divergences in the full theory~\cite{guidoc}. 
In the latter case the infrared divergences are automatically regularized 
by the non-perturbative contributions in the physical matrix elements and
no renormalization is required.
\par As discussed in ref.~\cite{ua},
it is not clear  whether  the use of $S_{\scriptscriptstyle LEET}(x)$ 
will be convenient in practice, since this propagator is much more singular than the
full one. For this reason we expect
that  the  correlation functions computed in numerical simulations
using $S_{\scriptscriptstyle LEET}(x)$ will be affected by larger 
statistical fluctuations.
\section*{Acknowlegments}
L.S. acknowledges the support
of German Bundesministerium f\"ur Bildung und Forschung under contract
06 TM 874 and DFG Project Li 519/2-2.
We acknowledge partial support by M.U.R.S.T.  


\begin{thebibliography}{99}
\bibitem{general}
I.I.~Balitsky, V.M.~Braun, A.V.~Kolesnichenko,
Nucl. Phys. B 312 (1989) 509;
V.M.~Braun, I.B.~Filyanov, Z.~Phys. C 44 (1989) 157; ibid.  48 (1990) 239;
V.L.~Chernyak, I.R.~Zhitnitsky, Nucl.~Phys. B345 (1990) 137;
P.~Ball, V.M.~Braun, H.G.~Dosch, Phys.~Rev. D 44 (1991) 3567.
\bibitem{ruckl} V.M.~Belyaev, A.~Khodjamirian, R.~R\"uckl,
Z. Phys. C60 (1993) 349;  A.~Khodjamirian and R.~R\"uckl 
WUE-ITP-97-049,  to appear in  Heavy Flavors, 2nd edition, eds., A.J.~Buras 
and M. Linder (World Scientific), hep-ph/9801443. 
\bibitem{cernyak} V.L.~Chernyak, A.R.~Zhitntsky, Phys.~Rep. 112 (1984) 173. 
\bibitem{ali} A.~Ali, V.M.~Braun and H.~Simma, Z.~Phys. C63 (1994) 437. 
\bibitem{ball} P.~Ball and  V.M.~Braun, Phys.~Rev. D54 (1996) 2182;
 P.~Ball, V.M.~Braun, Y.~Koike, K.~Tanaka, hep-ph/9802299. 
\bibitem{lanl:semilept-lat94} T.~Bhattacharya and R.~Gupta,
  Proc. Lattice 94, 12th Int. Symp. on Lattice Field Theory,
Bielefeld, Germany, 1994, \npbps{42} (1995) 935
and {47} (1996) 481
\bibitem{ape:hl-semilept} APE Collaboration, C.R.~Allton \etal,
  \plb{345} (1995) 513
\bibitem{elc:hl-semilept} As.~Abada \etal, \npb{416} (1994) 675
\bibitem{ukqcd:hlff} UKQCD Collaboration, D.R.~Burford \etal,
  \npb{447} (1995) 425
\bibitem{ukqcd:btorho} {UKQCD} Collaboration, J.M.~Flynn \etal,
  \npb{461} (1996) 327
\bibitem{ukqcd:hlfits} UKQCD Collaboration, L.~Del Debbio \etal,
  Granada--Marseille--South\-ampton
  preprint, UG--DFM--4/97,
  CPT--97/P.3505, SHEP--97--13, hep-lat/9708008.
\bibitem{sac1} G.~Martinelli and C.T.~Sachrajda,  Phys.~Lett. 190B (1987) 15 
and  Nucl.~Phys. B316 (1989) 305.  
\bibitem{smrenormalon} G.~Martinelli and C.T.~Sachrajda,
Phys.~Lett.~B354 (1995) 423 and  Nucl.~Phys.~B478 (1996) 660.
\bibitem{ua} U.~Aglietti et al., ROME1-1204/98, hep-ph/9804416.
\bibitem{grin} M.J.~Dugan and B.~Grinstein, Phys.~Lett. B255 (1991) 583.
\bibitem{koster} G.P.~Korchemsky and G.~Sterman, Phys.~Lett. B340 (1994) 96;
A.G.~Grozin and G.P.~Korchemsky,   Phys.~Rev. D53 (1996) 1378. 
\bibitem{guidoc} U.~Aglietti, G.~Corb\`o and L.~Trentadue, hep-ph/9712237.
\end{thebibliography}
\end{document}